\documentstyle[11pt,aaspp4]{article}

\def\mathnew{\mathsurround=0pt}
\def\simov#1#2{\lower .5pt\vbox{\baselineskip0pt \lineskip-.5pt
        \ialign{$\mathnew#1\hfil##\hfil$\crcr#2\crcr\sim\crcr}}}
\def\simg{\mathrel{\mathpalette\simov >}}
\def\siml{\mathrel{\mathpalette\simov <}}
\def\Mesz{M\'esz\'aros~}
\def\Pacz{Paczy\'nski~}
\def\beq{\begin{equation}}
\def\enq{\end{equation}}
\def\bea{\begin{eqnarray}}
\def\ena{\end{eqnarray}}
\def\E52{E_{52}}
\def\r13{r_{13}}
\def\et2{\eta_2}
\def\no{n_0}
\def\th2{\theta_j^2}
\def\Ent{ \E52 / \no \th2 }
\def\t5{t_5}
\def\kap3{\kappa_3}
\def\bec{\begin{center}}
\def\enc{\end{center}}
\def\Sign20{\Sigma_{n20}}

\begin{document}
\voffset=-0.5 in

%\bec Draft :~~{5/15/98} \enc
%\bec Draft :~~{\today} \enc

\title{The Edge of a Gamma Ray Burst Afterglow}

\author{P. \Mesz$^1$ \& M.J. Rees$^2$} 
\noindent
$^1$Dpt. of Astronomy \& Astrophysics, Pennsylvania State University,
University Park, PA 16803 \\
$^2$Institute of Astronomy, University of Cambridge, Madingley Road, Cambridge
CB3 0HA, U.K. 
\bec
{\it MNRAS, accepted 16 July 1998; submitted 4 June 1998}
\enc
%\maketitle

\begin{abstract}
We discuss the formation of spectral features in the decelerating ejecta of
gamma-ray bursts, including the possible effect of inhomogeneities. These
should lead to blueshifted and broadened absorption edges and resonant
features, especially from H and He. An external neutral ISM could produce
detectable H and He, as well as Fe X-ray absorption edges and lines.
Hypernova scenarios may be diagnosed by Fe K-$\alpha$ and H Ly-$\alpha$
emission lines.

\end{abstract}


\section{Introduction}

Gamma-ray bursts (GRB) have been localized through spectral line measurements 
of a presumed host galaxy in two cases so far (Metzger et al., 1997, Kulkarni, 
et al., 1998). The GRB environment and its effect on the detectability and 
spectral properties of the afterglow are the subject of debate (e.g. Livio,
et al., 1997; Van Paradijs, J., 1998); a dense environment may be more typical 
of a massive stellar progenitor (Paczy\'nski, 1997, Fryer \& Woosley, 1998) 
while medium to low 
density environments could suggest compact merger progenitors (e.g. Bloom, 
Sigurdsson \& Pols, 1998). The lack of an optical afterglow following a detected 
X-ray afterglow (e.g. GRB 970828) may be due to a steep temporal fall-off of the 
flux (Hurley, 1997) or may be connected with absorption in a dense gas around 
the GRB (Groot, et al., 1997). The detection of spectral 
signatures which can be associated with the GRB environment would be of great
interest both for distance measurements and for helping to answer the above
questions, while spectral features associated with the burst ejecta itself 
would provide information about the fireball dynamics and its chemical 
composition, and clues about the triggering mechanism and the progenitor system.
In this paper we investigate the possibility of detectable spectral features 
arising in the shocked gas and in dense inhomogeneities coexisting with it, 
during the decelerating external shock phase of the burst afterglow. We also
consider the absorption features arising from gas outside the region ionized 
by the energetic photons emitted by the burst, and the reprocessing of X-ray
and optical photons by the external environment, including possible
signatures for a hypernova scenario.

\section{Afterglow Continuum Radiation}

The common model of GRB afterglows considers that their radiation arises 
in the decelerating blast wave of fireball material, produced e.g. by a compact 
binary merger or stellar collapse, impacting on an external medium. The 
radius at which deceleration begins is
$ r_d= 10^{16.7} (\Ent )^{1/3} \et2^{-2/3} \hbox{cm}$,
at an observer time 
$ t_{d}= r_d/(2 c \eta^2 )= 10^{1.9} (\Ent )^{1/3} \et2^{-8/3} \hbox{s}$,
where $E_o=10^{52}\E52 \hbox{erg~s}^{-1}$ is the initial fireball energy, 
$\eta=10^2 \et2$ is the terminal coasting bulk Lorentz factor, $M_o=E/\eta c^2$ 
is the rest mass entrained in the initial fireball, and $n_{ext}=1~\no$ 
cm$^{-3}$ is the average external density. 
After the contact discontinuity starts to decelerate, a forward blast wave
advances into the external medium and a reverse shock moves into the fireball 
gas. A similar situation arises also in the scenario where the mass and 
energy injection is not a delta function at $E_o,~M_o$ but rather continues
for some time (much shorter than $t_{d}$), adding more mass and energy as the
bulk Lorentz factor decreases,
\beq
M(>r) \propto \Gamma^{-w}~~~,~~~ E(>r) \propto \Gamma^{-w+1}~,
\label{eq:refresh}
\enq
(Rees \& \Mesz, 1998) in which case the shock is continually reenergized 
(refreshed shocks) by the new energy and mass arriving at radius $r$ (the
reverse and forward shocks being assumed thin relative to $r$). The contact
discontinuity radial coordinate $r$ and its Lorentz factor $\Gamma$ vary as
$r  = 10^{17.5+q_1} (\Ent)^{2/h} \et2^{2(h-8)/h} \t5^{(h-6)/h}~\hbox{cm}$ and
$\Gamma  =  10^{0.84+q_2} (\Ent)^{1/h} \et2^{(h-8)/h} \t5^{-3/h}$.
Here $h=7+A+w$, with $h=8$ for the standard adiabatic impulsive model ($A=1,~
w=0$), and the correction factors $q_1=3.1[(h-6)/h-1/4]$, $q_2=9.3[(1/h - 1/8]$ 
are non-zero with $h \neq 8$ only for refreshed ($w\neq 0$) and/or radiative
($A=0$) cases.
%and one recovers $\Gamma \propto r^{-3/2} \propto t^{-3/8},~ r\propto t^{1/4}$ 
%while for a fully radiative remnant $A=0,~H=7$ and one recovers $\Gamma\propto 
%r^{-3} \propto t^{-3/7}$. 
The postshock comoving magnetic field, whose energy density is a fraction $\xi$ 
of the thermal proton energy, is
$B = 10^{-0.5+q_2} (\E52/\th2)^{1/h} \no^{(h-2)/2h} \xi_{-2}^{1/2} 
\et2^{(h-8)/h} \t5^{-3/h} ~\hbox{G}$, 
and the observer-frame synchrotron peak frequency is
\beq
\nu_m = 10^{14.32+4 q_2} (\E52/\th2)^{4/h} \no^{(h-8)/h} \et2^{4(h-8)/h} 
\xi_{-2}^{1/2} \kap3^2 \t5^{-12/h}~\hbox{Hz}.
\label{eq:num}
\enq
The comoving synchrotron cooling time for electrons radiating at the 
observed frequency $\nu_m$ is
$t_{sy}=10^{5.36 -3 q_2} (\E52/\th2)^{-3/h} \no^{(3-h)/h)} \xi_{-2}^{-1} 
\kap3^{-1} \et2^{-3(h-8)/h} \t5^{9/h}$ s, while the comoving expansion
time $t_{ex} = (r/c\Gamma) = 10^{5.86+q_1-q_2} (\Ent)^{1/h} \et2^{(h-8)/h} 
\t5^{(h-3)/h}$ s. The comoving inverse Compton time 
$t_{ic}\sim 0.3 \xi_{-2}^{1/2}\kap3^{1/2}t_{sy}$ is of the same order as
the synchrotron time, so its effect on the dynamics will be neglected.
The synchrotron efficiency is high until $\t5 \sim 1$ and drops afterwards as
$e_{sy} \sim (t_{ex}/t_{sy})$. 
%(For the nominal parameters taken here, $\mu= 4(\xi/\kappa)(m_p/m_e) \sim 
% 0.1 \kap3^{-1} \xi_{-2}$, the inverse Compton time remains a factor $\sim 0.3$
% shorter than the synchrotron time, and may be neglected for our purposes).
In the regime where the shortest timescale is the expansion time, the
comoving intensity $I'_{\nu'_m} \sim 4 \no \Gamma (\sigma_T c B^2 \kappa^2
\Gamma^2 /8\pi) c t_{ex}$, and for a source at luminosity distance $D$ the 
flux density at observer frequency $\nu_m$ is $F_{\nu_m} \simeq c^2 t^2 D^{-2} 
\Gamma^5 I'_{\nu'_m}$, or 
\beq
F_{\nu_m} = 10^{30.2+q_1+6q_2}~D^{-2} (\E52/\th2)^{8/h}\no^{(3h-16)/2h}
\et2^{8(h-8)/h} \xi_{-2}^{1/2} \t5^{3(h-8)/h} ~\hbox{erg/s/cm$^2$/Hz},
\label{eq:fnum}
\enq
while the source-frame luminosity is $L_{\nu_m}= 4\pi\theta_j^2 D^2 F_{\nu_m}$.
%L_{\nu_m} = 10^{31+q_1+6q_2}~ \E52^{8/H}\theta_j^{2(H-8)/H}\no^{(3H-16)/2H}
%\xi_{-2}^{1/2} \et2^{8(H-8)/H} \t5^{3(H-8)/H} .
%\label{eq:lnum}
The continuum flux at detector frequency $\nu_d$ for a spectrum of the form 
$\propto \nu^{\alpha}$ is $F_{\nu_d} \sim
%(\Mesz \& Rees, 1997)
F_{\nu_m} (\nu/\nu_d)^{\beta} \propto t^{\delta}$, decreasing as a power 
law in time after the peak passes through the detector band $\nu_d$ (where
$\delta$ depends on $\beta$ as well as on $W,~A$ and possibly other parameters,
e.g. Wijers, Rees \& \Mesz, 1997, Vietri, 1997, Katz \& Piran, 1997, Sari,
1997, \Mesz, Rees \& Wijers 1998, Rees \& \Mesz, 1998).

\section{ Absorption in the decelerating ejecta }

The   baryon density downstream of the forward blast wave and of the reverse 
shock, as long as the expansion remains relativistic, is dominated by the 
 ejecta, rather than by baryons swept up in the external shock. 
Their total number is $N_p(r)= N_{po} (\Gamma(r)/\Gamma_o)^{-w}$, where 
$N_{po}=(E_o/\eta m_p c^2)$ is the value for an impulsive fireball and $E_o$ 
and $\Gamma_o=\eta$ are the initial energy and Lorentz factor. The 
corresponding baryon column density $\Sigma_p=N_p/(4\pi \th2 r^2)$ is
\beq
\Sigma_p = 10^{16.8+q_3-2q_1} \E52^{(h-w-4)/h} \theta_j^{(w+4-2h)/h}
 \no^{(w+4)/h} \et2^{(8w-5h+32)/h} \t5^{(3w-2h+12)/h} \hbox{cm}^{-2},
\label{eq:Sigmap}
\enq
where $q_3=9.3w/h$.
The refreshed case ($w\neq 0$) is interesting because $\Sigma_p$ can be 
larger than in the impulsive case. The mean comoving baryon density downstream 
of the reverse shock is $n_p= \Sigma_p / c t_{ex}$ or
\beq
n_p = 10^{0.44-3q_1+q_2+q_3} \E52^{(h-w-5)/h} \theta_j^{(w+5-2h)/h} 
 \no^{(w+5)/h} \et2^{(8w-6h+40)/h} \t5^{3(w-h+5)/h} \hbox{cm}^{-3}~.
\label{eq:np}
\enq
For $\t5\siml 1$ the cooling time $t_{cool} < t_{ex}$, and the 
comoving electron inverse Compton temperature is
\beq
T_{ic}= h\nu_m/h\Gamma = 10^{3.1+3q_2} (\E52/\th2)^{3/h} \no^{(h-6)/h} 
 \et2^{3(h-8)/h} \xi_{-2}^{1/2} \kap3^2 \t5^{-9/h}~\hbox{K}, 
\label{eq:Tic}
\enq
while the black-body temperature is of order $T_{bb}\sim 10^{2.5-q_1/4+2q_2}$ K.
% \E52^{3/2H} \no^{(H-6)/4H} \theta_j^{-3/H} \et2^{3(H-8)/2H} \xi_{-2}^{1/4} 
%\kap3^{1/2} \t5^{(H-18)/4H}$ K. 
At the (adiabatic postshock) density (\ref{eq:np}) the recombination time 
$t_{rec}\sim 3\times 10^{11} T_3^{1/2} Z^{-2} n_p^{-1}$ s exceeds the comoving 
expansion time.  
However, if $t_{cool} < t_{ex}$ and there is good coupling between protons and electrons, then Compton cooling behind the shock affects the protons as well. 
The shock would then be radiative ($A=0$) and the gas adjusts in pressure 
equilibrium to a density 
\beq
n_{eq} \sim n_o m_p c^2 \Gamma / k T_{ic} \sim 10^{10.6 -2q_2} 
 (\E52/\th2)^{-2/h} \no^{(h+4)/2h} \et2^{-2(h-8)/h} \t5^{6/h} \hbox{cm}^{-3}. 
\label{eq:neq}
\enq
At this density (\ref{eq:neq}) the recombination time is much shorter than the 
expansion time, the ionization parameter in the comoving frame $\Xi= L_{\nu_m} 
\nu_m /n_b r^2 \Gamma^2 \siml 1$, and the hydrogen as well as heavier elements 
in the shocked ejecta will be in their neutral state (Kallman \& McCray, 1982).

The outflow may also include denser blobs or filaments of thermal material, 
entrained from the surrounding debris torus or condensed through instabilities 
in the later stages of the outflow. Previously (\Mesz \& Rees, 1998) we 
considered the effect of such blobs in internal shocks at $r\sim 10^{13}$ cm 
leading to $\gamma$-ray emission. Here we consider the effects of blobs that 
catch up with external shocks from the initial part of the ejecta around $r\sim 
10^{16}-10^{17}$ cm. One cannot predict how much material would be present in
such blobs, but possible instabilities affecting them would be minimized when 
their Lorentz factor $\Gamma_b$ is close to that of the surrounding flow. 
For an equipartition magnetic field $B$ in the flow frame $\Gamma$, the pressure 
equilibrium blob density at the inverse Compton temperature in its comoving 
frame is
\beq
n_b \sim (B^2/8\pi k T_{ic})(\Gamma_b/\Gamma) = 10^{12.1} (\Gamma_b/\Gamma)
 \no^{1/2}\kap3^{-2}~\hbox{cm}^{-3}.
\label{eq:nb}
\enq
If the smoothed-out blob density seen in the flow frame is taken to be a
fraction $\alpha$ of the mean flow density, ${\bar n_b} = \alpha n_p 
(\Gamma/\Gamma_b)$, the smoothed-out baryon column density in blobs is
\beq
{\bar \Sigma_b} = \alpha \Sigma_p (\Gamma/\Gamma_b)^2 ~,
\label{eq:Sigmab}
\enq
c.f. equation (\ref{eq:Sigmap}). The filling factor of such blobs, 
$f_v=({\bar n_b}/n_b)$ is very small, of order $10^{-11.7-3q_1+q_2+q_3}$,
%10^{-11.7-3q_1+q_2+q_3} \alpha (\Gamma/\Gamma_b)^2 \E52^{(H-W-5)/H} 
%\no^{(2W-H+10)/2H} \theta_j^{(W-2H+5)/H} \kap3^2 \et2^{(8W-6H+40)/H} 
%\t5^{3(W-H+5)/H}$
while the surface covering factor $f_s={\bar \Sigma_b}/(n_b r_b)$ can be
larger than unity for blob radii smaller than $r_b=\alpha \Sigma_p n_b^{-1} 
(\Gamma/\Gamma_b)^2$, which is of order $\sim 10^{5 -2q_2+q_3}$ cm, for
nominal parameters taken at $\t5 \sim 1$.
%r_b \sim 10^{4.7 -2q_2+q_3} \alpha f_s^{-1} (\Gamma/\Gamma_b) \E52^{(H-W-4)/H} 
 %\no^{(2W-H+8)/2H} \theta_j^{(W-2H+4)/H} \et2^{(8W-5H+32)/H} \kap3^2
 %\t5^{(3W-2H+12)/H} \hbox{cm}.
%\label{eq:rb}
At these blob densities and temperatures the recombination time is very short
compared to expansion times, and also the ionization parameter in the blob
comoving frame $\Xi_b= L_{\nu_m} \nu_m /n_b r^2 \Gamma_b^2 \leq 1$, so
hydrogen, and also heavier elements inside the blobs, will be in their 
neutral state. 

For hydrogenic atoms the absorption cross section at threshold is 
$\sigma_a = 7.9\times 10^{-18} Z^{-2}$ where $Z$ is the effective ionic charge. 
If the fraction of ions with charge $Z$ is $x_z$, the opacity at the ionization 
edge is 
\beq
\tau_Z \sim 0.8 x_z Z^{-2} \Sigma_{17} \propto \t5^{~(-2h+12+3w)/h} ~,
\label{eq:tauz}
\enq
where the baryon column density $\Sigma_{17}=(\Sigma /10^{17}\hbox{cm}^{-2})$
may be due either to the neutral diffuse gas, if e-p coupling is effective 
before the adiabatic stage (equation [\ref{eq:Sigmap}]), or to blobs and 
filaments (equation [\ref{eq:Sigmab}]) (absorption from blobs would occur
for blob velocities $\Gamma_b \geq \Gamma(t)$). Note that for $\t5 <1$ or for 
continued input ($w\neq 0$, $h>8$) $\Sigma$ could be larger than $10^{17}$ 
cm$^{-2}$.  For H, the observed frequency of 
the edge would be at $13.6 \Gamma$ eV, with $\Gamma$ given below
equation (1), or around $0.1 (1+z)^{-1}$ KeV at $\t5 \sim 1$. For HeII, 
whose ionization edge is 54.4 eV at rest, the observed edge is around $0.4 
(1+z)^{-1}$ KeV in the observer frame. If the blobs are made up predominantly
of heavy metals, e.g. Fe, the rest-frame edge is near 9.2 KeV, or 
\beq
h \nu_Z \sim 64 ~(Z/26)^2 (1+z)^{-1} 10^{q_2} (\Ent)^{1/h} \et2^{(h-8)/h} 
  \t5^{-3/h}~\hbox{KeV}
\label{eq:nuz}
\enq
in the observer frame, for ions of charge $Z$ in a GRB at redshift $z$. 
The edges will generally not be sharp, since they will be observed from a 
ring-like region around the edges of the front hemisphere of the remnant 
(Panaitescu \& \Mesz, 1998), over which the simultaneously observed radiation 
samples a bulk Lorentz factor range of at least $\Delta\Gamma/\Gamma \sim 
\Delta \nu/\nu \sim 0.3$. The time dependences of equations (\ref{eq:tauz},
\ref{eq:nuz}) refer to the diffuse gas and also to blobs, provided the density 
(or $\alpha$ parameter) of the latter is appreciable over the range of values 
$\Gamma_b \geq \Gamma(t)$.  Resonant lines from Ly-$\alpha$ lines of H and He 
have cross sections comparable to those for ionization, and would be expected 
at energies redwards of the absorption edges. They absorb over a narrow energy 
range and will therefore produce only a shallow and wide trough due to the 
above Lorentz factor smearing. On the other hand, H and He emission lines from 
recombination in the dense cooled ejecta or in blobs should be more prominent, 
even when broadened  by $\Delta\nu/\nu \sim 0.3$, since they correspond to a 
much larger amount of energy taken out from the continuum bluewards of the 
absorption edges. Such broad emission features would enhance the detectability
of the drop seen just bluewards of it from the continuum absorption.

\section{Absorption in an External Neutral Medium}

The X-ray and UV photons from GRB will ionize the surrounding medium out to a 
radius which can be estimated from the total number of ionizing photons 
produced. For simplicity, we assume in this section a canonical adiabatic  
impulsive afterglow ($h=8,~A=1,~w=0$). From (\ref{eq:fnum}) the synchrotron 
peak luminosity is $L_{\nu_m}=4\pi D^2 F_{\nu_m} \sim 10^{31} \E52 \no^{1/2} 
\xi_{-2}^{1/2}$ erg s$^{-1}$ Hz$^{-1}$, and the time when $\nu_m$ reaches 
13.6 eV is (equation [\ref{eq:num}]) $t_{13.6} \sim 1.6 \times 10^4 (\E52/
\theta_j^2)^{1/3} \xi_{-2}^{3/4} \kap3^3$ s. The total number of ionizing 
photons produced  is $N_i \sim L_{\nu_m}~ t_{13.6} / h \sim 2.4\times 10^{61} 
\E52^{4/3} \theta_j^{-2/3} \no^{1/2} \xi_{-2}^{5/4} \kap3^2$. Loeb \& Perna 
(1998) have calculated the time dependence of the equivalent widths of atomic 
lines for an afterglow flux time dependence $\propto t^{-3/4}$, for which the 
largest (integrated) contribution to the ionization happens at late times,
and show that for finite cloud sizes of column density $\Sigma_p\sim 3\times 
10^{20}\hbox{cm}^{-2}$ the equivalent widths would vary considerably over 
timescales of days to weeks. Recent afterglows indicate that the more commonly
observed continuum flux time dependences are steeper than $t^{-1}$, so that
most of the ionizing photons are 
created at early times, within the first few hours. In this case, after an 
initial transient similar to that described by Loeb \& Perna (1998),  one expects the edges and equivalent widths to stabilize; 
we consider their behavior after this time, but before recombination occurs 
(e.g. few hours $\siml  t \siml$ year). 

The ISM may generally extend beyond 
the finite ionized region considered by Loeb \& Perna (1998), in which case 
the ionization structure would be photon-bounded, rather than density-bounded, 
i.e. there is neutral matter  beyond the ionization zone.  The ionization 
radius is 
\beq
R_i \sim ( N_i /4 \no \theta_j^2 )^{1/3} \simeq 
  2\times 10^{20} \E52^{4/9} \no^{-1/6} \xi_{-2}^{5/12} \kap3 ~\hbox{cm}~.
\label{eq:Ri}
\enq
The $\no^{-1/6}$ weak dependence is model-specific (equation [\ref{eq:fnum}]),
and could be $\no^{-1/3}$ in a more generic source.  For a typical (neutral)
column density $\Sigma_n=10^{20}\Sign20$ cm$^{-2}$ (beyond $R_i$, but within 
the galaxy, e.g. the neutral component of a galactic disk) one gets a K edge 
optical depth 
\beq
\tau_Z \sim 0.8\times 10^3 x_z Z^{-2} \Sigma_{n20}~,
\label{eq:tauzext}
\enq
where $x_z$ is the fraction of the species with effective nuclear charge $Z$.
This is large for H and He, while if $\Sigma_n\sim 2\times 10^{21}$ (for which 
the local visual absorption $A_v=5\times 10^{-22} \Sigma_n$ would reach one 
magnitude) the optical depth of Fe at solar abundances ($x_{Fe,\odot}= 3 
\times 10^{-4}$) would be $\tau_{Fe} \sim 0.1$, and similarly for other metals.
Unlike the blue-shifted ejecta edges of equation (\ref{eq:nuz}), the ISM
K-edge observed energy is
\beq
h \nu_Z \sim 13.6 ~Z^2 (1+z)^{-1} ~\hbox{eV}~\sim ~
 9.2 ~(Z/26)^2 (1+z)^{-1}~\hbox{KeV}~,
\label{eq:nuzext}
\enq
bluewards of which the flux is blanketed up to $\nu/\nu_z \sim 10~ Z^{-2/3} 
(x_z \Sign20 )^{1/3}$.  

Resonant  Ly-$\alpha$ absorption  from hydrogen in the neutral ISM will 
be conspicuous for high enough column densities. The Doppler broadened 
Ly-$\alpha$ would have a large optical depth at line center, the equivalent 
widths being  dominated by the damping wings, in the square-root regime of the 
curve of growth, $(W_\nu / \nu ) = ( r_e \lambda_{lu}^2 c^{-1} f_{lu} A_{ul} 
\Sigma)^{1/2}$. For analogous hydrogen-like K-$\alpha$ resonant 
transitions in other species,
\beq
(W_\nu / \nu ) = 8.3\times 10^{-13} (x_z \Sigma_n)^{1/2} \simeq
 10^{-2} x_z^{1/2} {\Sign20}^{1/2}~.
\label{eq:eqw}
\enq
For $\Sign20 \simg 1$ one gets H  equivalent widths of order tens of 
Angstroms. Similar features in He, and in other elements like C, would be blocked out by HI continuum absorption at the same frequencies. But this absorption would not affect Fe, and the 6.7 keV X-ray Fe K-$\alpha$ widths at solar abundances are 
of order tens of eV.
Other possibly-detectable features include the Fe edge at 9.1 keV and the O VIII edge at 0.871 keV.  Some Si lines 
at 1.66 keV and 2.28 keV may also be detectable. For continuum fluxes $\propto t^{-1.2}$ 
or steeper, these should remain constant after the first few hours.

X-ray photons can also be re-emitted by Fe fluorescent inner-shell scattering 
in the external ISM, and much interest has been raised by the possibility of 
X-ray and UV/O emission lines from this. The source X-ray continuum can be 
approximated as a pulse of radial width $c\delta t \sim 10^{13} \delta t_3$ cm 
(for X-ray light curves decaying faster than $t^{-1}$). This X-ray pulse 
occupies the volume between the two paraboloids given by the equal-arrival time 
surfaces $r(1-\cos\theta)=ct$ and the same for $c(t+\delta t)$, where $t$ is 
observer time, $r$ is distance from source center and $\theta$ is the polar 
angle variable (we ignore here effects from a possible jet opening angle 
$\theta_j$).  The base of the paraboloid towards the observer is the ionization 
radius $R_i$ (equation [\ref{eq:Ri}]), and the medium outside the outer 
paraboloid and/or $R_i$ is neutral. The total number of X-ray continuum photons 
emitted over $4\pi$ is $N_{x} \sim L_{\nu_m} h^{-1} \delta t \sim 10^{60}
\E52^{3/2} \no^{1/2}\xi_2 \kap3^2$ photons. The fraction incident on the 
paraboloid is roughly $N_x \theta^2 \sim 10^{53} t_3 R_{i20}^{-1}$ ph ($\theta 
\ll 1$, thin paraboloids). The optical depth of the paraboloid wall is $\tau_f 
\sim \no x_{Fe} \sigma_f \delta r \sim \no x_{Fe} \sigma_f R_i \delta t/t \sim 
6\times 10^{-4} x_{-3.5} \no R_{i20}\delta t_3  t_3^{-1} $, where $\sigma_f 
\sim 10^{-20}\hbox{cm}^2$. The number of fluorescent X-rays is 
$N_f \sim N_x \theta^2 \tau_f \sim 3\times 10^{49} x_{-3.5}\no \delta t_3$ 
and the luminosity $L_f\sim  3\times 10^{38} x_{-3.5} \no$ erg/s gives a flux 
$F_{f}\sim 3\times 10^{-19}\E52^{3/2} \no^{3/2} x_{-3.5} \xi_{-2} \kap3^2 
D_{28}^{-2}$ erg cm$^{-2}$ s$^{-1} \sim$ constant. A similar estimate is 
obtained using the photons scattered by the ``back" (near) half of the 
paraboloid. The light pulse volume between the paraboloids is $\delta V =
(\pi/3)c^3 t^3 [(1+\delta t /t)^3-1] [(R_i /ct)^2-1] \sim \pi R_i^2 c\delta t$, 
constant for $t \gg \delta t$. An upper limit on the fluorescent 
photons that can be produced is $N_{f,m} \siml \delta V \no Z x_{Fe} \sim 
3\times 10^{52} \no x_{-3.5} R_{i,20}^2 (Z/26) \delta  t_3$ ph, each Fe 
scattering 26 photons before being fully ionized (recombinations take too long 
at ISM densities; this is also why H and He emission lines from the ISM 
outside $R_i$ are negligible). With a timescale $10^3 \delta t_3$ s the 
luminosity is $L_{f,m}\sim 3\times 10^{41}$ erg/s, and the flux upper limit is 
$F_{f,m}\sim 3\times 10^{-16} \E52^{4/9} \no^{5/6} x_{-3.5} \xi_{-2}^{5/12} 
\kap3 D_{28}^{-2}$ erg cm$^{-2}$ s$^{-1}$. The continuum is $F_x \sim 
10^{-10} \E52 \no^{1/2} \xi_{-2}^{1/2} D_{28}^{-2}$ erg cm$^{-2}$ KeV$^{-1}$ 
s$^{-1}$, and the equivalent width is $W_\nu/\nu \siml 10^{-6} \E52^{-5/9} 
\no^{1/3} x_{-3.5} \xi_2^{-1/12} \kap3$. Fluorescent emission lines from the
ISM would thus be hard to detect, unless the host galaxy ISM environment is 
exceptionally dense (leading to high optical extinction), or metal-rich.
If detected, they would offer useful information on the nature of (and location
within) the host galaxy.

\section{ Possible Hypernova Signatures}

In contrast to the previous discussion, where ISM conditions were assumed,,
the lines and edges may be more prominent in a hypernova scenario, because
the circumburst environment could be much denser.  Whereas NS-NS or NS-BH 
mergers can lead to a BH + torus system producing magnetic fireballs of 
$10^{53}-10^{54}$ ergs (e.g. \Mesz \& Rees, 1997; Narayan, \Pacz \& Piran, 
1992), in the hypernova scenario a similar system and energy is derived from 
the collapse of a single or binary fast rotating star, or a He-BH merger 
(Paczy\'nski, 1997, 1998; Fryer \& Woosley, 1998), i.e. it involves a close 
stellar companion and/or a massive envelope. The extent of the dispersed medium
should depend on whether there has been a radiation-driven (slow) outflow before
the burst, such as expected if the event were preceded by inward spiralling of a  white dwarf or BH
through a companion's atmosphere. Consider, as an example, a dispersed envelope 
of $1 M_\odot$ or $N_a \sim 10^{57}$ nucleons spread over a radius 
$3 \times 10^{15} r_{15.5}$ cm. Its
density is $n\sim 10^{10} M_{0} r_{15.5}^{-3}$ cm$^{-3}$, with a Thompson depth 
$\tau_T \sim$ few. The fireball propagates in a less dense funnel along the
rotational axis, and for $\no \sim 10^2 n_{f2},~\et2\simg 1$ the deceleration 
radius and X/O afterglow radius are inside the envelope. At observer times $t 
\sim 10^5\t5$ s most of the afterglow photons are optical, and the envelope 
electron inverse Compton temperature is $T\sim 10^4$ K. The ionization parameter
$\Xi \sim L/n r^2 \sim 10^6$ so Fe is mostly Fe XXVI. The recombination time is 
$t_{rec}\sim 5\times 10^2 T_4^{1/2} Z^{-2} n_{10}^{-1}$ s, which is 
$\sim$ 1 s for Fe. Each Fe ion can reprocess $t/t_{rec}\sim 10^5 \t5$ 
continuum X-ray photons into lines, and the total number of Fe recombinations 
is $N_{rec,Fe} \sim N_a x_{Fe}(t/t_{rec}) \sim 10^{59} M_0 x_{-3} T_4^{-1/2} 
n_{10}$.  The number of continuum X-ray photons at $t\sim 10^5$ s is $N_x\sim 
10^{61}\E52 n_{f2}^{1/2}$ ph. A fraction $\sim 10^{-2} \E52^{-1} n_{f2}^{-1/2} 
M_0 x_{-3} n_{10}$ of the X-ray continuum can thus be reprocessed into Fe 
lines, which is significant.  For H, the recombination time is $\sim 
5\times 10^2$ s, and the number of H recombinations at time $t$ is $N_{rec,H} 
\sim N_a (t/t_{rec})\sim 2\times 10^{60} M_0 T_4^{-1/2} n_{10}$, while the 
number of optical continuum photons is $\sim 10^{63}\E52 n_{f2}^{1/2}$,
so H recombinations can also produce a Ly-$\alpha$ flux significant compared 
to the continuum flux.  The equivalent widths are $W_\nu/\nu \sim 10^{-2}$
for Fe K-$\alpha$ and $W_\nu/\nu \sim 10^{-3}$ for H Ly-$\alpha$.
Besides these emission lines, absorption edges may also be seen if the 
observer line of sight goes through the envelope.  

A variant of this scenario occurs when the envelope is more massive
and is Thompson optically thick, e.g. there is a funnel created by a
puffed up companion or a common envelope, with the GRB at the
center. In this case, when the beaming angle $\Gamma^{-1}$ is wider
than the funnel, a substantial fraction of the emitted X-ray and
optical continuum would be reflected from the funnel walls. By analogy
with AGN reflection models (e.g. Ross and Fabian, 1993) one would
expect detectable Fe edges and K-$\alpha$ features (at 9.1 and 6.7 keV
respectively), as well as a Ly edge and Ly-$\alpha$ features imprinted
in the reflected optical component.

\section{Discussion}

We have shown that the decelerating external shock of a fireball afterglow 
may produce  a significant absorption edge in the cooled shocked ejecta,
as long as cooling is faster than adiabatic losses and protons are well
coupled to electrons. Absorption edges can also arise from cool, dense
blobs or filaments in pressure equilibrium with the shocked smooth ejecta.
These edges can reach optical depths of order unity and will be 
blue-shifted by a mean bulk Lorentz factor of typically $\Gamma\sim 5-7$ 
around $t\sim 1$ day, with larger optical depths and blue-shifts at earlier
times. The edges will be broadened by the spread $\Delta\nu/\nu \sim 
\Delta\Gamma/\Gamma \sim 0.3$ in the ring of observed material. The H, He 
and Fe edges would be at energies $\sim ~(0.1,~0.4,~9.2) (1+z)^{-1} 
\t5^{-3/8}$ KeV for a standard adiabatic remnant around $t\sim 10^5\t5$ s or 
1 day (equation [\ref{eq:nuz}]). Strong blueshifted Fe edges would only be
expected in the presence of metal-rich blobs, whereas H and He edges could
arise in the diffuse cooled ejecta. The latter would be a diagnostic for the 
$A=0$ radiative dynamical regime (which, e.g., for a $w=0$ no-injection case 
in a homogeneous external medium evolves as $\Gamma\propto r^{-3}$). 

Information on the dynamics of the explosion  may be obtained from the time 
dependence of the edge characteristics. For instance, it may be possible to 
distinguish between an impulsive adiabatic case ($A=1$, $w=0$, $h=8$), an 
impulsive radiative case ($A=0$, $w=0$, $h=7$) and a refreshed shock case 
(say $A=1$, $w=2$, $h=10$ as an example) because the edge depth (equation 
[\ref{eq:tauz}]) would vary as $t^{-1/2}$, $t^{-2/7}$ or $t^{-2/10}$, 
while the edge energies (equation [\ref{eq:nuz}]) would vary as $t^{-3/8},~
t^{-3/7},~t^{-3/10}$, the third set of numbers being for this particular $w$.

The energetic photons will ionize the surrounding matter out to a radius $R_i
\sim 60 \E52^{4/9}\no^{-1/6}$ pc (equation [\ref{eq:Ri}]), and after a brief
initial transient lasting less than a few hours, reprocessing would be 
negligible except in the external environment. The neutral gas outside this 
ionized region will produce the absorption edges and resonant absorption lines
typical of the ISM in the host galaxy. These would be non-blueshifted 
absorption features, unaffected by any broadening from bulk Lorentz factor 
smearing in the afterglow. They should be affected only by a cosmological 
redshift, and would thus provide valuable distance information. The H Lyman 
continuum optical depths can be very substantial for modest galactic disk 
column densities, as are the Ly$-\alpha$ absorption equivalent widths 
(equations [\ref{eq:tauzext},\ref{eq:eqw}]), which would be in the optical 
range for $z\simg 3$.  Reprocessed emission lines from the ISM would be hard
to detect, unless the event occurs in an exceptionally dense or metal-rich
environment. If detected, they would offer useful information on the nature 
of, and location within, the host galaxy.

A hypernova scenario could be distinguished by the presence of a significant
flux of Fe K-$\alpha$ and H Ly-$\alpha$ emission lines, reprocessed by a
moderately Thompson optically thick companion or envelope. For a more massive,
Thompson optically thick envelope, a significant reflected component would be 
expected, in which Fe absorption edges and K-$\alpha$ features, as well as 
hydrogen Lyman edge and Ly-$\alpha$ features would be imprinted.

\acknowledgements
{This research has been supported by NASA NAG5-2857 and
the Royal Society. We are grateful to B. \Pacz, G. Garmire and members of
the Swift team for stimulating comments}

\end{document}